\documentclass[10pt]{iopart}

\usepackage{graphicx}
\usepackage{dcolumn}
\usepackage{bm}
\usepackage{color}

\begin{document}

\title[CFQS AE]{Study of the Alfven Eigenmodes stability in CFQS plasma using a Landau closure model}


\author{J. Varela}
\ead{jacobo.varela@nifs.ac.jp}
\address{Universidad Carlos III de Madrid, 28911 Leganes, Madrid, Spain}
\address{National Institute for Fusion Science, National Institute of Natural Science, Toki, 509-5292, Japan}
\author{A. Shimizu}
\address{National Institute for Fusion Science, National Institute of Natural Science, Toki, 509-5292, Japan}
\author{D. A. Spong}
\address{Oak Ridge National Laboratory, Oak Ridge, Tennessee 37831-8071, USA}
\author{L. Garcia}
\address{Universidad Carlos III de Madrid, 28911 Leganes, Madrid, Spain}
\author{Y. Ghai}
\address{Oak Ridge National Laboratory, Oak Ridge, Tennessee 37831-8071}

\date{\today}

\begin{abstract}
The aim of this study is to analyze the stability of the Alfven Eigenmodes (AE) in the Chinese First Quasi-Axisymmetric Stellarator (CFQS). The AE stability is calculated using the code FAR3d that solves the reduced MHD equations to describe the linear evolution of the poloidal flux and the toroidal component of the vorticity in a full 3D system, coupled with equations of density and parallel velocity moment for the energetic particles (EP) species including the effect of the helical couplings and acoustic modes. The Landau damping and resonant destabilization effects are added in the model by a given closure relation. The simulation results indicate the destabilization of $n=1$ to $4$ AEs by EP during the slowing down process, particularly $n=1$ and $n=2$ Toroidal AEs (TAE), $n=3$ Elliptical AE (EAE) and $n=4$ Non circular AE (NAE). If the resonance is caused by EPs with an energy above $17$ keV (weakly thermalized EP), $n=2$ EAEs and $n=3$ NAEs are unstable. On the other hand, EPs with an energy below $17$ keV (late thermalization stage) lead to the destabilization of $n=3$ and $n=4$ TAEs. The simulations for an off-axis NBI injection indicate the further destabilization of $n=2$ to $4$ AEs although the growth rate of the $n=1$ AEs slightly decreases, so no clear optimization trend with respect to the NBI deposition region is identified. In addition, $n=2,4$ Helical AE (HAE) are unstable above an EP $\beta$ threshold. Also, if the thermal $\beta$ of the simulation increases (higher thermal plasma density) the AE stability of the plasma improves. The simulations including the effect of the Finite Larmor Radius (FLR) and electron-ion Landau damping show the stabilization of the $n=1$ to $4$ EAE/NAEs as well as a decrease of the growth rate and frequency of the $n=1$ to $4$ BAE/TAEs.
\end{abstract}

%
%
%
%
%


\vspace{2pc}

\maketitle

\ioptwocol

\section{Introduction \label{sec:introduction}}

The analysis of the plasma stability in fusion devices with configuration designs that test different types of quasi-symmetry is essential for the design and optimization of a future Stellarator fusion reactor. One promising example is the Chinese First Quasi-Axisymmetric Stellarator (CFQS), a low aspect ratio device designed to combine the positive features of Tokamaks and Stellarators. CFQS targets the operation of plasma with a high $\beta$, good neoclasical  transport and improved MHD stability \cite{1,2,3,4,5,6,7}.

The CFQS device as well as the National Compact Stellarator Experiment (NCSX) \cite{8} are based on quasi-axisymmetry although there are other possibilities, for example Quasi Poloidal Stellarator (QPS) applying quasi-poloidal symmetry \cite{9} or the Helically Symmetric Experiment (HSX) using quasi-helical symmetry \cite{10}. In addition, there are generalized symmetries such as omnigenity where the mean radial collisionless guiding center magnetic drift is minimized, leading to good collisionless orbit confinement \cite{11}. The optimization of AE stability in these configurations is important for efficient plasma heating, and to reduce the operational requirements of a future nuclear fusion reactor and for improved economic viability.

The CFQS plasma will be heated by a tangential neutral beam injector (NBI) with an injection energy of $30$ keV and a power of $0.9$ MW \cite{12}. The energetic particles (EP) generated by the tangential NBI can drive instabilities and enhance EP transport, decreasing the heating efficiency of a CFQS plasma. The effect of the EP driven instabilities was already observed in other devices such as TFTR, JET and DIII-D tokamaks or LHD, TJ-II and W7-AS stellarators \cite{13,14,15,16,17,18,19,20,21,22}. The instability is triggered if the mode frequency resonates with the drift, bounce or transit frequencies of the EP.

Alfv\' en Eigenmodes (AE) exist in the spectral gaps of the shear Alfv\' en continua \cite{23,24}. There are different Alfv\' en eigenmode families linked to frequency gaps produced by periodic variations of the Alfv\' en speed, for example ($n$ is the toroidal mode and $m$ the poloidal mode): toroidicity induced AE (TAE) couple $m$ with $m+1$ modes \cite{25,26,27}, beta induced AE driven by compressibility effects (BAE) \cite{28}, ellipticity induced AE (EAE) coupling $m$ with $m+2$ modes \cite{29,30}, noncircularity induced AE (NAE) coupling $m$ with $m+3$ or higher \cite{31,32} and helical AE (HAE) coupling $n$ with $n+L$ ($L$ is the device magnetic field period) \cite{33,34} and $m$ with $m+\Delta m$ where $\Delta m$ is an integer. 

The present study shows the first systematic analysis of the AE stability in CFQS plasmas, identifying the AEs that could be triggered by an EP component. The destabilization threshold is calculated for $n=1$ to $4$ AEs triggered by EP with energies between $10$ to $30$ keV, similar to the energy of the EP injected by a tangential NBI during the slowing down process before the thermalization. In addition, the effect of the NBI deposition region, finite thermal $\beta$ and helical couplings on the AE stability are studied. Also, some optimization trends with respect to the AE stability are suggested regarding variations in the fast ion density profile. Finally, the effect of the Finite Larmor Radius (FLR) and electron-ion Landau damping on the AE stability is analyzed. This study is part of global project dedicated to analyze the plasma stability of devices with different symmetries, identifying the configuration that shows the best plasma stability with respect to AE modes as well as possible optimization trends regarding the NBI operational regime, thermal plasma properties and magnetic configuration.

The gyro-fluid code FAR3d is used to perform this study \cite{35}, which is an extended version of the original FAR3d code that solves the reduced linear resistive MHD equations \cite{36,37,38}, adding the moment equations of the energetic ion density and parallel velocity \cite{39,40} reproducing the linear wave-particle resonance effects required for Landau damping/growth. The simulations are based on equilibria calculated by VMEC code \cite{41}. The main advantage of the gyro-fluid code FAR3d is the computational efficiency based on the reduction of selected kinetic effects to a set of 3D fluid-like equations rather than more complex approaches. On the other hand, the simplification of the kinetic effects can lead in some cases to a deviation of FAR3d results compared to more complete approaches, although a methodology has been developed for calibrating the Landau-closure against more complete kinetic models through optimization of the closure coefficients \cite{42}. Also, the FAR3d code assumes a Maxwellian distribution for the EP which has the same second moment, effective EP temperature, with respect to the slowing down distribution. A parametric analysis with respect to the EP energy and $\beta$ is performed approximating the resonances triggered by a slowing down distribution function by a set of Maxwellian distribution functions. Please see the appendix for further information. In addition, a benchmarking study performed between hybrid and gyro-kinetic codes and FAR3d calculated a similar growth rate, frequency and mode structure for reverse shear and toroidal AE in DIII-D plasma heated by tangential NBIs \cite{43}. It should be noted that present study may be understood as a first step in the analysis of the AE activity in NBI heated CFQS plasma. The study conclusions should be bench-marked with simulations performed by kinetic or gyro-kinetic codes, so that the Landau closure and FAR3d code calibration is validated. Nevertheless, the Landau closure used in this analysis was already successfully employed in other stellarators as the LHD, TJ-II and Heliotron J.

This paper is organized as follows. The model equations, numerical scheme and equilibrium properties are described in section \ref{sec:model}. The EP $\beta$ threshold of $n=1$ to $4$ AEs for different EP energies is analyzed in section \ref{sec:stability}. The effect of the NBI deposition region on the AE stability is studied in section \ref{sec:deposition}. The stability of helical AEs is calculated in section \ref{sec:helical}. The effect of the thermal $\beta$ on the AE stability is analyzed in section \ref{sec:thermal}. Next, the effect of the FLR and e-i Landau damping is studied in section \ref{sec:damping}. Finally, the conclusions of this paper are presented in section \ref{sec:conclusions}.

\section{Numerical model \label{sec:model}}

The FAR3d code solves the linear evolution of the thermal plasma (poloidal flux, total pressure, toroidal component of the vorticity and thermal parallel velocity) coupled with the equations of the EP density and parallel velocity moments. The numerical model uses finite differences in the radial direction and Fourier expansions in the angular variables for the equilibrium flux coordinates ($\rho$, $\theta$, $\zeta$). A semi-implicit initial value solver is used to resolve the numerical scheme. The present model was already used to study the AE stability in DIII-D, ITER, LHD, TJ-II and Heliotron J \cite{44,45,46,47,48} as well as the EIC stability in LHD \cite{49}, indicating reasonable agreement with the observations. The reader can find more details of the numerical modes in these references.

A single EP Maxwellian distribution cannot reproduce the same resonance with respect to a slowing down distribution, because the drive of the AE modes is determined by the gradient of the phase space distribution. Thus, simulations using an anisotropic slowing down distribution function are required to confirm the instabilities calculated by FAR3d code. Nevertheless, the analysis includes a parametric analysis with respect to the EP energy and $\beta$, that is to say, a set of Maxwellian distribution functions are used to approximate the resonances triggered by a slowing down distribution function. This information is useful for optimization studies. It should be noted that the model reproduces the destabilizing effect of the passing EP, although the effect of anisotropic beams or ICRF driven EP cannot be modeled by the present version of the code. However, the pitch angle of the EP generated by the tangential NBI in CFQS plasma should be small thus the model approximation is valid.

\subsection{Equilibrium properties}

\begin{figure}[h]
\centering
\includegraphics[width=0.45\textwidth]{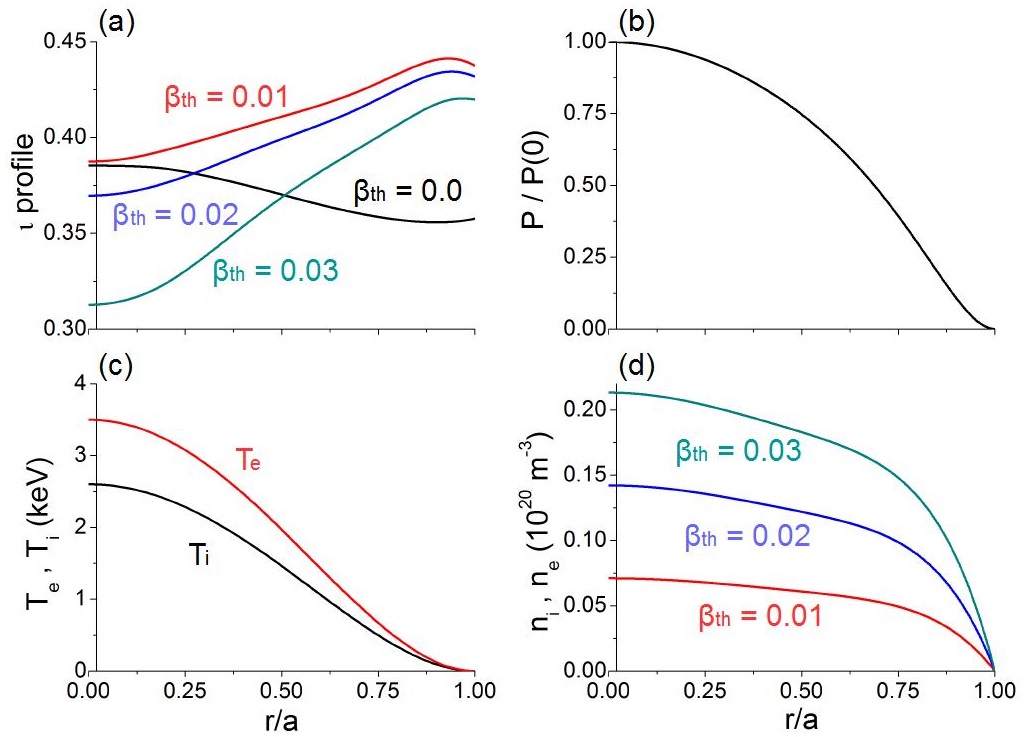}
\caption{(a) $\rlap{-} \iota$ profile, (b) total pressure, (c) thermal plasma temperature and (d) thermal plasma density.}
\label{FIG:1}
\end{figure}

\begin{figure}[h]
\centering
\includegraphics[width=0.45\textwidth]{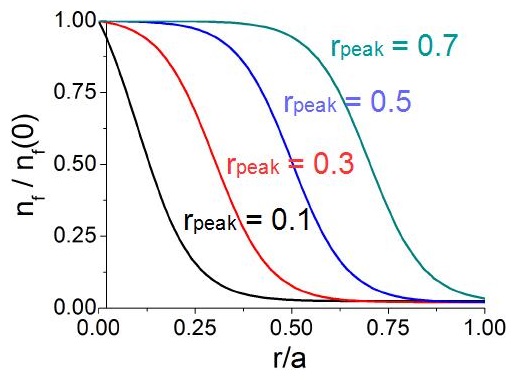} 
\caption{EP density profiles used in the study.}
\label{FIG:2}
\end{figure}

A set of free boundary results from the VMEC equilibrium code \cite{50} is calculated for the vacuum CFQS configuration ($\beta_{th}=0$) and finite thermal $\beta$ cases with $\beta_{th}=0.01$, $0.02$ and $0.03$. The effect of the bootstrap current is included in the equilibria, calculated using the code BOOTSJ \cite{51}. The magnetic field at the magnetic axis is $1$ T, the toroidal field period number is $2$, the averaged inverse aspect ratio is $\varepsilon=0.25$ and the major radius $R_{0} = 1$ m. The energy of the injected particles by the NBI is $T_{f}(0) = 30$ keV and the nominal energy of the EP resulting in an averaged Maxwellian energy equal to the average energy of a slowing-down distribution is $17$ keV. Figure~\ref{FIG:1} shows the main profiles of the thermal plasma. The thermal $\beta$ of the model increases because the thermal plasma density increases. Figure~\ref{FIG:2} shows the EP density profiles used in the study. The energy of the EP is assumed constant, with no radial dependency, for simplicity. It should be noted that the EP density profile and energy is the same in the simulations for the vacuum configuration and finite $\beta_{th}$. The EP $\beta$ ($\beta_{f}$) is defined at the magnetic axis. The growth rate ($\gamma \tau_{A0}$) is normalized to  the Alfv\' en time at the magnetic axis ($\tau_{A0} = R_0 (\mu_0 \rho_m)^{1/2} / B_0$).

\begin{figure}[h]
\centering
\includegraphics[width=0.45\textwidth]{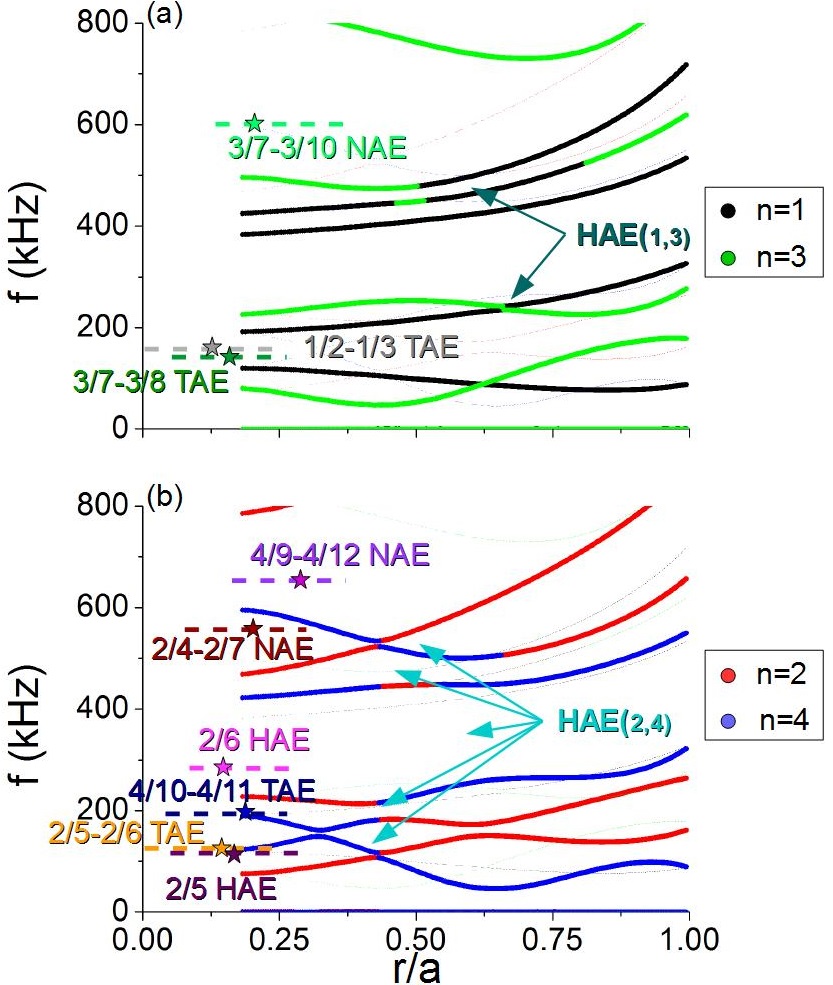} 
\caption{Alfven continuum of the vacuum CFQS configuration for the $n=1$ to $4$ modes.}
\label{FIG:3}
\end{figure}

Figure~\ref{FIG:3} shows the Alfv\' en gaps of the vacuum CFQS configuration for the $n=1$ to $4$ toroidal families including the effect of the helical couplings. The Alfv\' en gaps are calculated by the code Stellgap adding the effect of the sound wave \cite{52}. The sound wave spectrum is simplified by using the 'slow sound' approximation \cite{53}. This approximations retains the BAE gap although suppresses most of the lower frequency BAAE gap structure. The upper frequency range of the BAE gap is around $100$ kHz, the TAE gap up to $250$ kHz, the EAE gap up to $400$ kHz and the NAE gaps at higher frequencies. The frequency range of the gaps slightly changes between toroidal mode families showing also a radial dependency. In addition, there are several helical gaps caused by the coupling of the helical family $n=1,3$ at $230$ and $450$ kHz and by the helical family $2,4$ at $175$, $200$, $300$, $450$ and $500$ kHz. The $n=1-3$ helical gaps are narrower compared to the $n=2,4$ helical gaps. It should be noted that Stellgap simulations for the vacuum case are performed using a reference thermal ion density of $10^{20}$ m$^{-3}$ and a thermal electron temperature of $3.5$ keV; for this reason the BAE gap is observed, although the equilibrium for balance is for $\beta_{th} = 0$.

\subsection{Simulations parameters}

The simulations are performed using a uniform radial grid of 1000 points. The dynamic toroidal modes ($n$) in the simulation range from $n=1$ to $4$. The poloidal mode selection covers all the resonant rational surfaces (table~\ref{Table:1}). It should be noted that the equilibrium modes of the $n=2$ family are only included in the simulations with helical couplings. In the following, the mode number notation is $n/m$ consistent with the $\rlap{-}\iota=n/m$ definition.

\begin{table}[h]
\centering
\begin{tabular}{c}
Dynamic modes
\end{tabular}
\centering
\begin{tabular}{c | c}
\hline
$n$ & $m$  \\ \hline
$1$ & $[1,4]$  \\
$2$ & $[2,8]$  \\
$3$ & $[3,12]$  \\
$4$ & $[4,16]$  \\ \hline
\end{tabular}
\centering
\begin{tabular}{c}
Equilibrium modes
\end{tabular}
\centering
\begin{tabular}{c | c}
\hline
$n$ & $m$  \\ \hline
$0$ & $[0,10]$ \\
$2$ & $[0,4]$ \\ \hline
\end{tabular}

\caption{Dynamic and equilibrium toroidal and poloidal modes in the simulations with toroidal and helical couplings} \label{Table:1}
\end{table}

The kinetic closure moment equations (6) and (7) break the usual MHD parities, thus both parities $sin(m\theta + n\zeta)$ and $cos(m\theta + n\zeta)$ for all dynamic variables should be included in the simulation. The convention of the code is, in case of the pressure eigenfunction, that $n > 0$ corresponds to the Fourier component $\cos(m\theta + n\zeta)$ and $n < 0$ to $\sin(-m\theta - n\zeta)$. For example, the Fourier component for mode $-5/2$ is $\cos(-5\theta + 2\zeta)$ and for the mode $5/-2$ is $\sin(-5\theta + 2\zeta)$. The magnetic Lundquist number is assumed $S=5\cdot 10^6$.

Two different sets of simulations are performed, with and without helical couplings. First, the stability of each toroidal mode family is studied independently, thus only the toroidal couplings are included. Next, the effect of the helical couplings is added analyzing the joint evolution of the n=1,3 and n=2,4 helical families. In this way, the effect of the helical couplings in the AE growth rate and frequency is calculated. The analysis is limited to the dominant AE, that is to say, the mode with the largest growth rate; this is the mode that should most strongly limit the device performance.

\section{AE stability threshold \label{sec:stability}}

\begin{figure}[h]
\centering
\includegraphics[width=0.45\textwidth]{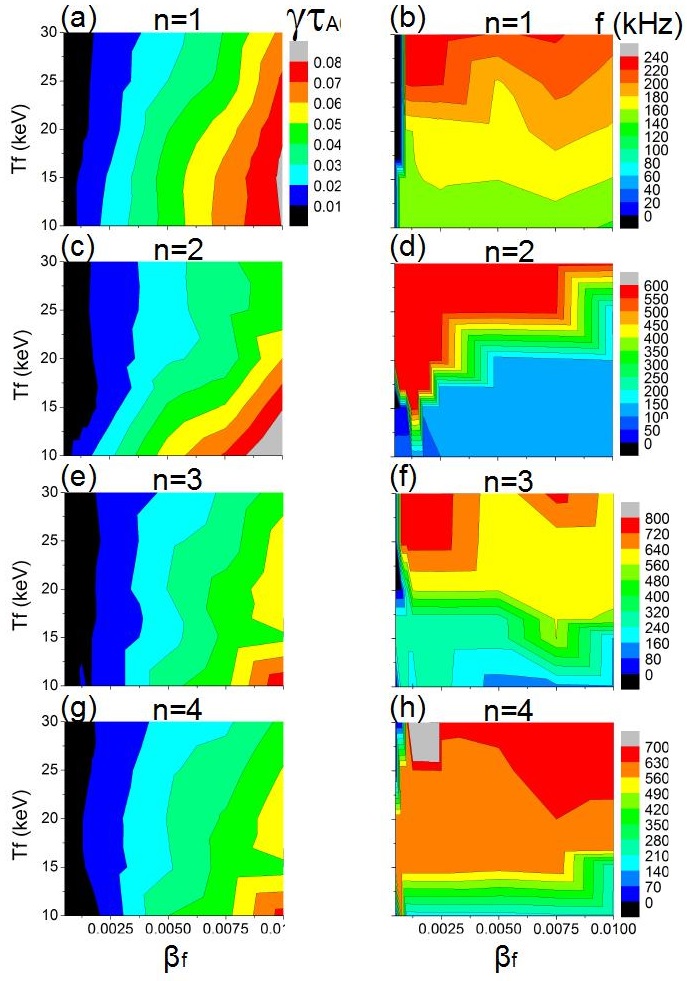} 
\caption{Growth rate and frequency of the $n=1$ to $4$ modes for different EP $\beta$ and energies.}
\label{FIG:4}
\end{figure}

Figure~\ref{FIG:4} shows the growth rate and frequency of the $n=1$ to $4$ toroidal families for different EP $\beta$ and energies. The AE growth rate increases with the EP $\beta$ because the EP density is higher, and hence the EP destabilizing effect (analogous to a higher NBI injection power). On the other hand, the AE growth rate decreases as the EP energy increases at fixed EP $\beta$ (EP density decreases proportionally to the increase of the EP energy). The simulations also show that the EP with $T_{f} \ge 15$ keV can trigger high frequency $n=2$ to $4$ AE (EAEs and NAEs). On the other hand, the $n=1$ AEs are destabilized in the frequency range of the BAE and TAE gaps.

\begin{figure}[h]
\centering
\includegraphics[width=0.45\textwidth]{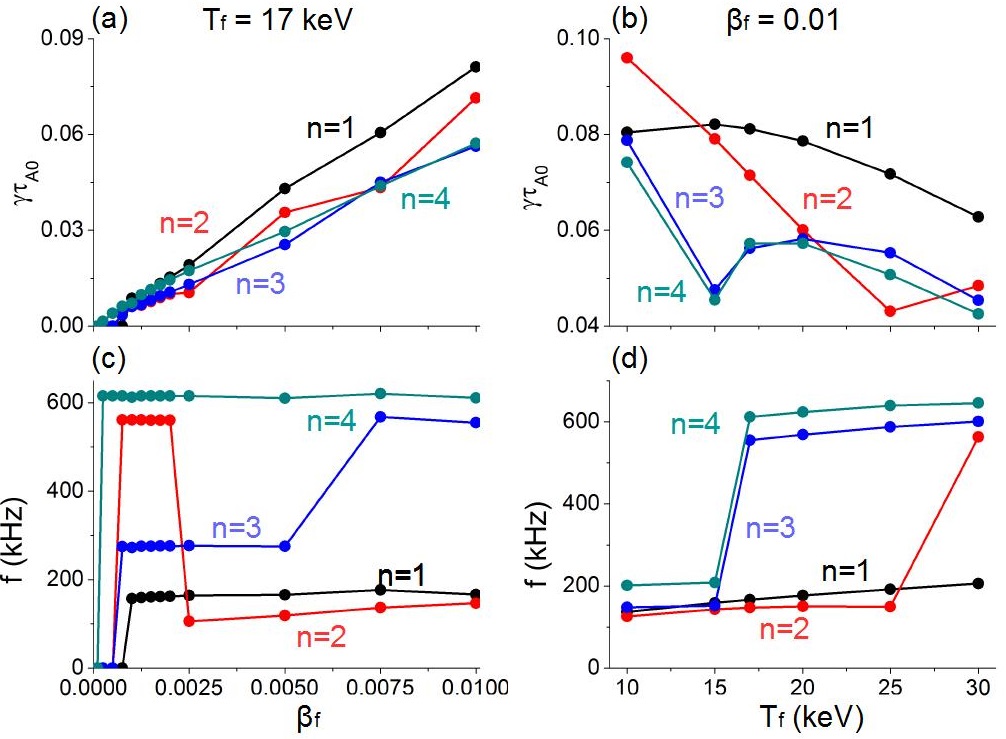} 
\caption{(a) Growth rate  and (c) frequency of the $n=1$ to $4$ modes for different EP $\beta$ ($T_{f}=17$ keV). (b) Growth rate and (d) frequency of the $n=1$ to $4$ modes for different EP energies (EP $\beta = 0.01$).}
\label{FIG:5}
\end{figure}

To clarify the stability trends of the AE with respect to the EP $\beta$ and energy, figure~\ref{FIG:5} shows the growth rate and frequency of the $n=1$ to $4$ AEs if the EP $\beta$ is scanned for an EP energy of $17$ keV (panels a and c) and if the EP energy is scanned for an EP $\beta=0.01$ (panels b and d). The simulations with a fixed EP energy indicate the destabilization of an $n=1$ AE with $f = 157$ kHz, $n=2$ AE with $f=562$ kHz, $n=3$ AE with $f=274$ kHz and $n=4$ AE with $f=616$ kHz if the EP $\beta$ increases above a given threshold. In addition, the $n=2$ AE shows a transition to a lower frequency AE family with $f = 106$ kHz and the $n=3$ AE to a higher frequency AE family with $f = 568$ kHz as the EP $\beta$ increases. The transition takes place because the growth rate of an AE at a different frequency range is larger, that is to say, there is a change of the dominant mode between configurations. Nevertheless, a transition does not necessarily mean the stabilization of the mode, the mode could be sub-dominant. The simulations with a fixed EP $\beta$ indicate a decrease of the $n=1$ to $4$ AEs growth rate as the EP energy increases. It should be noted that the AE growth rate increases if an AE of a higher frequency family is triggered, transition observed if $T_{f} \ge 17$ keV for the $n=3$ and $4$ AEs as well as if $T_{f} = 30$ keV for the $n=2$ AE.

\begin{figure}[h]
\centering
\includegraphics[width=0.45\textwidth]{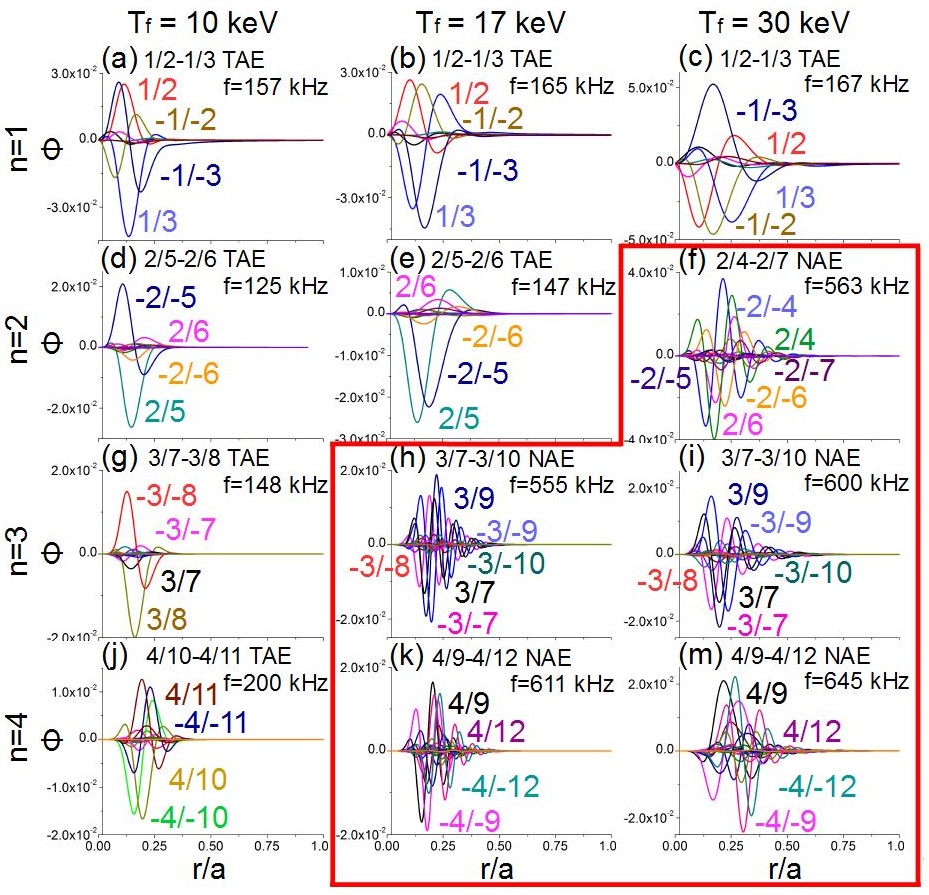} 
\caption{Eigenfunction of the $n=1$ AE if (a) $T_{f}=10$ keV, (b) $17$ keV and (c) $30$ keV for EP $\beta = 0.01$. Eigenfunction of the $n=2$ AE if (d) $T_{f}=10$ keV, (e) $17$ keV and (f) $30$ keV. Eigenfunction of the $n=3$ AE if (g) $T_{f}=10$ keV, (h) $17$ keV and (i) $30$ keV. Eigenfunction of the $n=4$ AE if (j) $T_{f}=10$ keV, (k) $17$ keV and (m) $30$ keV. The red box indicates the transition to a higher frequency AE family. Simulations EP $\beta = 0.01$.}
\label{FIG:6}
\end{figure}

Figure~\ref{FIG:6} shows the eigenfunction of the $n=1$ to $4$ AEs for different EP energies (EP $\beta = 0.01$). An $n=1$ TAE with $f \approx 160 $ kHz is unstable for all the EP energies analyzed, that is to say, during the EP slowing down process the resonances lead to the destabilization of a $1/2-1/3$ TAE (panels a to c). On the other hand, weakly thermalized EP populations ($T_{f} = 30$ keV), can destabilize an $2/5-2/7$ NAE with $f=563$ kHz, an $3/7-3/10$ NAE with $f=600$ kHz and an $4/9-4/12$ NAE with $f=645$ kHz (panels f, i and m). The red box indicates the eigenfunction of the high frequency AEs destabilized as $T_{f}$ increases. If $T_{f} = 10$ keV (EP at the end of the slowing down process) $n=2$ to $4$ TAEs are unstable (panels d, g and j). The $3/7-3/10$ and $4/9-4/12$ NAEs are also unstable if $T_{f} = 17$ keV (EP particles during the slowing down process, panels h and k). The frequency range of the AEs calculated are consistent with the frequency bands of the respective Alfv\' en gaps shown in fig~\ref{FIG:2}. It should be noted that the code shows in some simulations convergence issues leading to an artificial displacement of the local peak of the eigenfunction with coupled modes. Also, if the AE is destabilized near the upper/lower frequency range of a gap, the coupled modes can show different amplitudes as well as spurious low amplitude modes.

\section{Effect of the NBI deposition region on the AE stability \label{sec:deposition}}

In this section the stability of the AEs is studied with respect to the NBI deposition region, comparing on-axis and off-axis NBI injections.

\begin{figure}[h]
\centering
\includegraphics[width=0.45\textwidth]{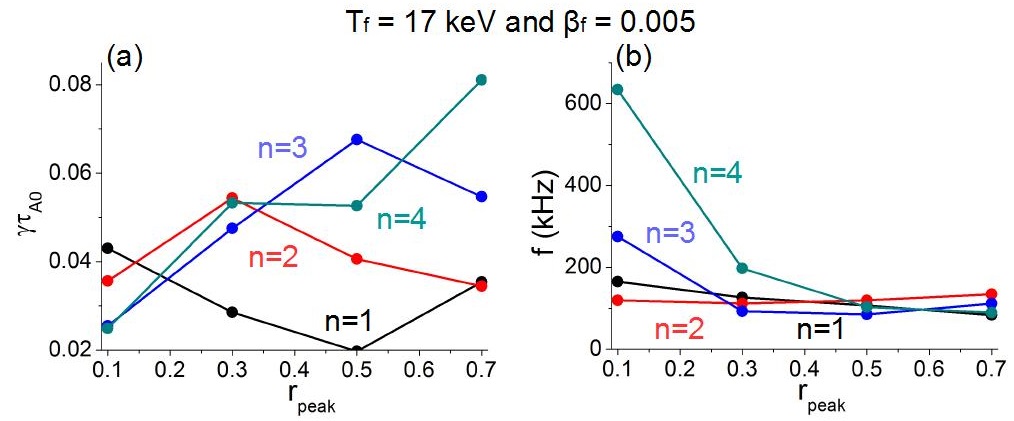} 
\caption{(a) Growth rate and (b) frequency of the $n=1$ to $4$ modes for different $r_{peak}$ values (EP $\beta = 0.005$ and $T_{f} = 17$ keV).}
\label{FIG:7}
\end{figure}

Figure~\ref{FIG:7} shows the growth rate (panel a) of frequency (panel b) of the $n=1$ to $4$ AEs if the location of the EP density gradient ($r_{peak}$) changes from $0.1$ to $0.7$ (EP $\beta = 0.005$ and $T_{f} = 17$ keV). If the on-axis ($r_{peak} = 0.1$) and off-axis ($r_{peak} \ge 0.1$) simulations are compared, the growth rate of the $n=1$ AEs decreases around a $50 \%$ if the NBI is deposited in the middle plasma region, although the growth rate of the $n=2$ to $4$ AEs increases. In addition, the off-axis NBI injection causes a transition to a lower frequency AE family of the $n=1$, $3$ and $4$ AEs if $r_{peak} \ge 0.3$. It should be noted that the dominant AE changes as the location of the drive is modified, because the gradient of the EP density profile is located at different radial location where the continuum structure is different.

\begin{figure}[h]
\centering
\includegraphics[width=0.45\textwidth]{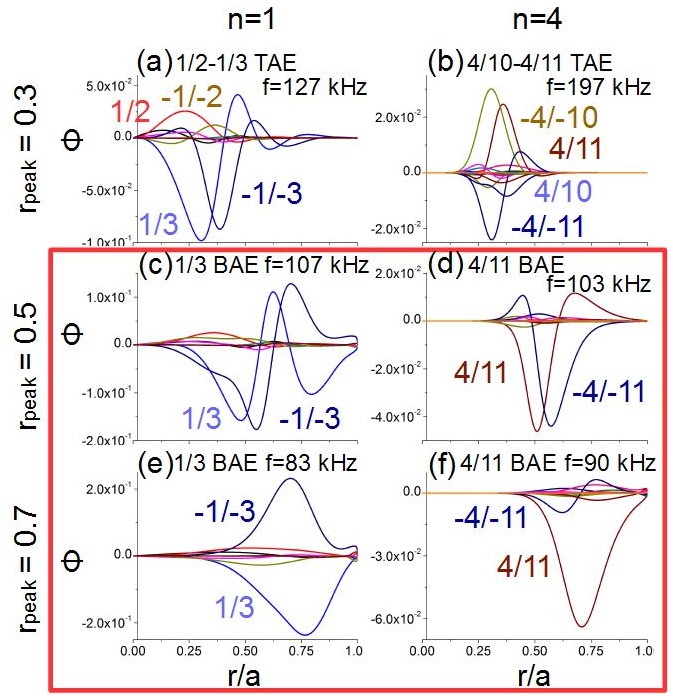} 
\caption{Eigenfunction of the $n=1$ AE if (a) $r_{peak} = 0.3$ keV (c) $0.5$ and (e) $0.7$. Eigenfunction of the $n=4$ AE if (b) $r_{peak} = 0.3$ keV (d) $0.5$ and (f) $0.7$. Simulations EP $\beta = 0.005$ and $T_{f}=17$ keV.}
\label{FIG:8}
\end{figure}

Figure~\ref{FIG:8} shows the eigenfunction of the $n=1$ and $4$ AEs as the NBI deposition region is located further off-axis. There is a transition from a $1/2-1/3$ TAE with $f=127$ kHz to a $1/3$ BAE with $f=107$ kHz if $r_{peak} = 0.5$ ($83$ kHz if $r_{peak} = 0.7$). The same transition is observed from a $4/10-4/11$ TAE with $f=197$ kHz to a $4/11$ BAE with $f=103$ kHz if $r_{peak} = 0.5$ ($90$ kHz if $r_{peak} = 0.7$).

\section{Effect of the helical couplings on the AE stability \label{sec:helical}}

This section is dedicated to analyze the stability of the AEs if the effect of the helical couplings is included on the model. The EP energy and EP $\beta$ required to destabilize helical AEs (HAE) of the helical families $n=1,3$ and $n=2,4$ is calculated.

\begin{figure}[h]
\centering
\includegraphics[width=0.45\textwidth]{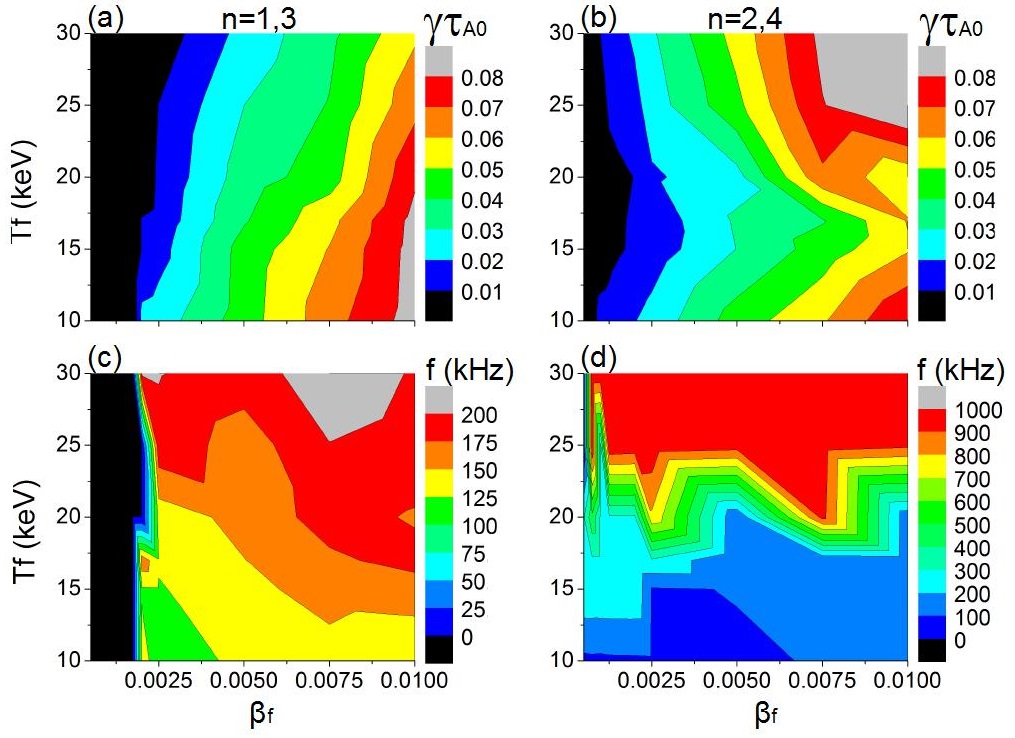} 
\caption{Growth rate and frequency of the $n=1,3$ and $2,4$ AEs for different EP $\beta$ and energies.}
\label{FIG:9}
\end{figure}

Figure~\ref{FIG:9} shows the growth rate and frequency of the $n=1,3$ and $2,4$ helical families for different EP $\beta$ and energies. The $n=1,3$ AEs are unstable if EP $\beta \ge 0.002$ (panels a and c). The $n=2,4$ AEs are unstable if EP $\beta \ge 0.00025$ (panels b and d). If $T_{f} \ge 20$ keV, the $n=2,4$ AE are destabilized in a frequency range above $900$ kHz, showing a transition of a higher frequency AE family.

\begin{figure}[h]
\centering
\includegraphics[width=0.45\textwidth]{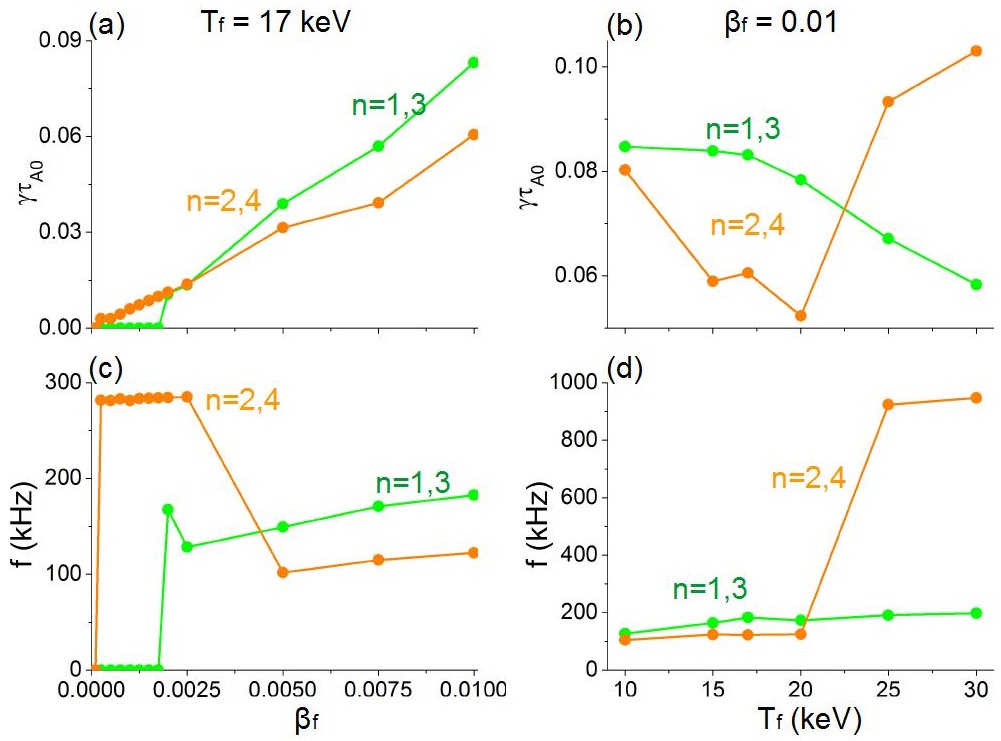} 
\caption{(a) Growth rate and (c) frequency of the $n=1,3$ and $2,4$ AEs for different EP $\beta$ ($T_{f} = 17$ keV). (b) Growth rate and (d) frequency of the $n=1,3$ and $2,4$ AEs for different EP energies (EP $\beta = 0.01$).}
\label{FIG:10}
\end{figure}

Figure~\ref{FIG:10} shows the growth rate and frequency of the $n=1,3$ and $2,4$ AEs for different EP $\beta$ ($T_{f}=17$ keV, panels a and c) and for different EP energies (EP $\beta=0.01$, panels b and d). An $n=1,3$ AE is triggered if EP $\beta = 0.002$ although the destabilization threshold is smaller for the $n=2,4$ AE (panel a). In addition, the $n=2,4$ AE shows a transition to a lower frequency AE as the EP $\beta$ increases (panel c). The  growth rate of the $n=1,3$ and $2,4$ AEs decreases as the EP energy increases, although if $T_{f} \ge 25$ keV the $2,4$ AE shows a transition to a higher frequency AE family and the mode growth rate increases up to a $30\%$, transition not observed for the $n=1,3$ AE.

\begin{figure}[h]
\centering
\includegraphics[width=0.45\textwidth]{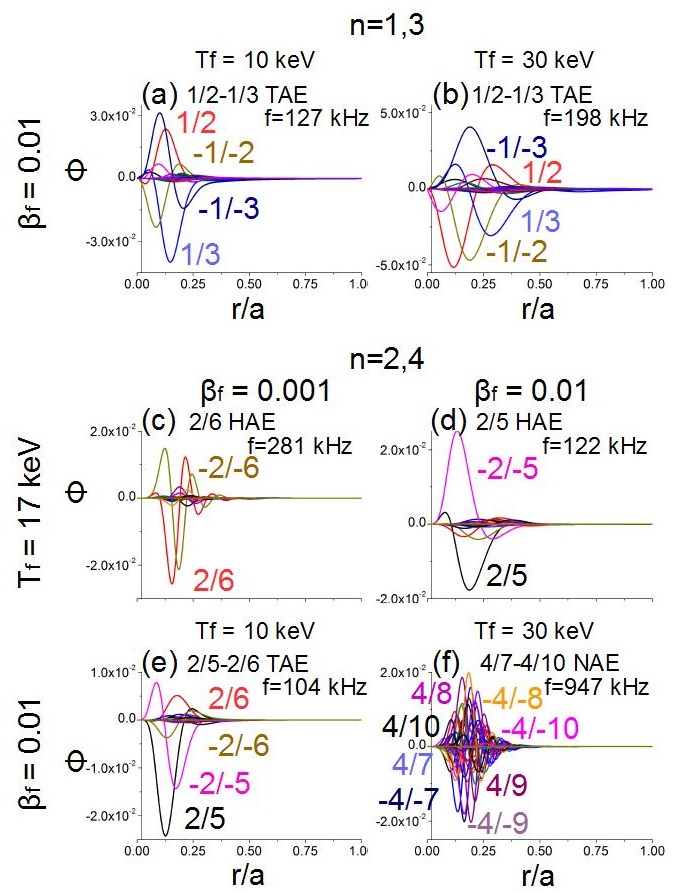} 
\caption{Eigenfunction of the $n=1,3$ AE if EP $\beta = 0.01$ and (a) $T_{f}=10$ keV or (b) $30$ keV. Eigenfunction of the $n=2,4$ AE if $T_{f}=17$ keV and (c) EP $\beta = 0.001$ or (d) $0.01$. Eigenfunction of the $n=2,4$ AE if EP $\beta = 0.01$ and (e) $T_{f} = 10$ keV and (f) $30$ keV.}
\label{FIG:11}
\end{figure}

Figure~\ref{FIG:11} shows the eigenfunction of the $n=1,3$ and $2,4$ AEs for different EP energies and EP $\beta$. The $n=1,3$ AE eigenfunction and growth rate is similar to the $n=1$ TAE (panels a and b), indicating that the helical couplings between the modes $n=1$ and $3$ are weak. In addition, the $n=1,3$ HAE gaps are very narrow at frequencies below $400$ kHz (see fig~\ref{FIG:2}), thus the $ n=1,3$ HAEs should be stable. On the other hand, the $n=2,4$ AE eigenfunction is different compared to the $n=2$ TAE and $n=4$ NAE if $T_{f} > 10$ keV, pointing out that the helical couplings between the modes $n=2$ and $4$ are strong enough to destabilize a $2/6$ HAE with $f=281$ kHz if EP $\beta = 0.001$ (panel c) and a $2/5$ HAE with $f=122$ kHz if EP $\beta = 0.01$ (panel d). It should be noted that the eigenfunction and frequency of the $n=2,4$ AE shows some differences compared to the $n=2$ TAE if $T_{f} = 10$ keV and the $n=4$ NAE if $T_{f} = 30$ keV, although the effect of the helical couplings is not large enough to destabilize a HAE.

\section{Finite thermal $\beta$ effect \label{sec:thermal}}

\begin{figure}[h]
\centering
\includegraphics[width=0.45\textwidth]{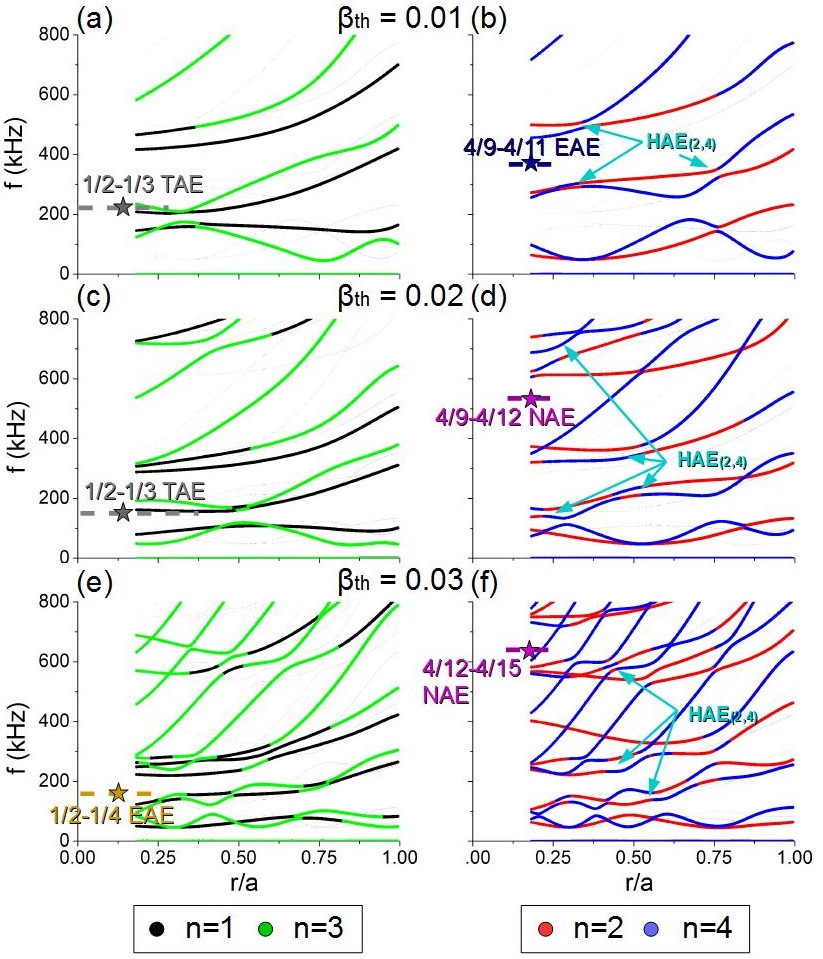} 
\caption{Alfven continuum for the CFQS configurations with a thermal $\beta$ of (a) $0.01$, (b) $0.02$ and (c) $0.03$.}
\label{FIG:12}
\end{figure}

This section is dedicated to study the finite thermal $\beta$ effect on the stability of the $n=1$ to $4$ AEs. The model $\beta_{th}$ increases because the thermal plasma density increases. The increase of $\beta_{th}$ causes a modification of the $\rlap{-} \iota$ profile, the plasma Alfven velocity and the Alfv\' en gap distribution, thus the AE stability also changes. Figure~\ref{FIG:12} shows the Alfv\' en gaps of CFQS configurations with $\beta_{th} = 0.01$, $0.02$ and $0.03$ for the $n=1$ to $4$ modes (helical couplings included). A higher $\beta_{th}$ leads to narrow width Alfv\' en gaps although a larger gap density. Consequently, the number of AEs that can be destabilized increases, yet these modes are more radially localized. Nevertheless, the range of frequencies and radial locations showing an enhanced continuum damping is larger as the thermal $\beta$ increases. In addition, the thermal ion FLR damping effect increases if the thermal $\beta$ increases, such us is discussed in the next section of the present document. In summary, the  AE stability may improve in high $\beta_{th}$ operations. It should be noted that the gaps of the helical family $n=1,3$ are very narrow, barely observed if $\beta_{th} = 0.01$. On the other hand, the gaps of the $n=2,4$ helical family can still be distinguished if $\beta_{th} = 0.03$.

\begin{figure}[h]
\centering
\includegraphics[width=0.45\textwidth]{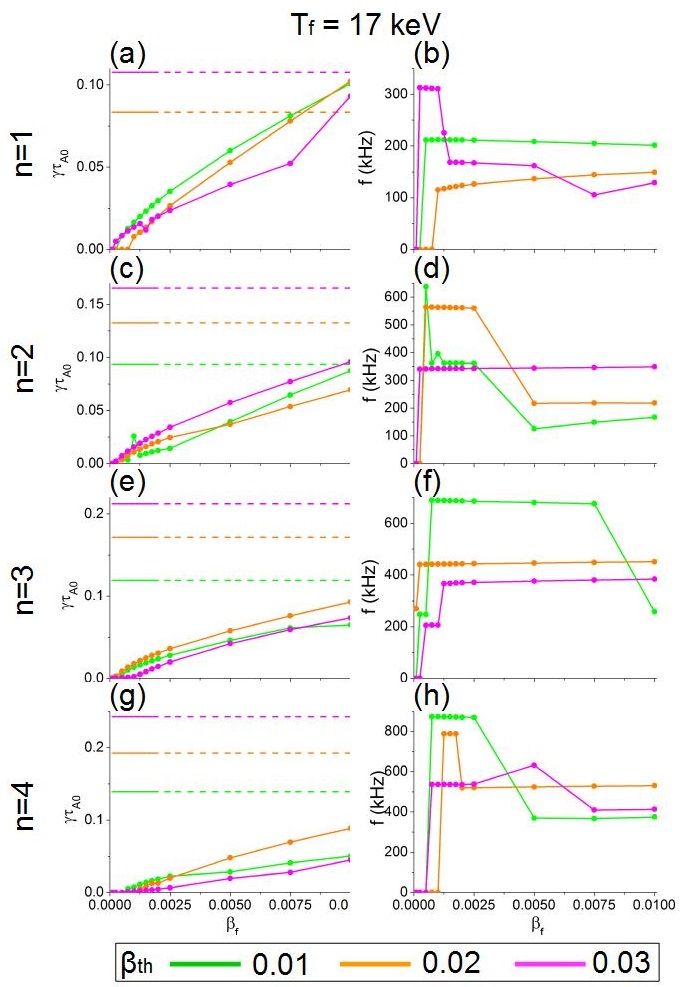} 
\caption{Growth rate and frequency of the $n=1$ to $4$ modes for different EP $\beta$ ($T_{f}=17$ keV) if $\beta_{th}=0.01$, $0.02$ and $0.03$. The dashed lines indicate the growth rate of the $n=1$ to $4$ pressure gradient driven modes.}
\label{FIG:13}
\end{figure}

Figure~\ref{FIG:13} shows the growth rate and frequency of the $n=1$ to $4$ AEs in simulations with different EP $\beta$ ($T_{f}=17$ keV) if $\beta_{th} = 0.01$, $0.02$ and $0.03$. The growth rate of the $n=1$ AE decreases if the $\beta_{th}$ increases from $0.01$ to $0.03$ (panel a), unless there is a transition to a higher frequency AE family causing an increase of the AE growth rate, for example the transition from a TAE to an EAE if $\beta_{th} = 0.03$ (panel b). The largest growth rate of the $n=2$ AE is observed in the simulations with $\beta_{th} = 0.03$ (panels c and d), although the $n=3$ and $n=4$ AEs with the largest growth rate are triggered if $\beta_{th} = 0.02$ (panels e and h). It should be noted that AEs in the frequency range of the TAE gap ($f < 250$ kHz) are only triggered by the $n=1$ to $3$ modes for a given range of EP $\beta$ and $\beta_{th}$, while the rest of the simulations show the destabilization of AEs in the frequency range of the EAE and NAE gaps.

\begin{figure}[h]
\centering
\includegraphics[width=0.45\textwidth]{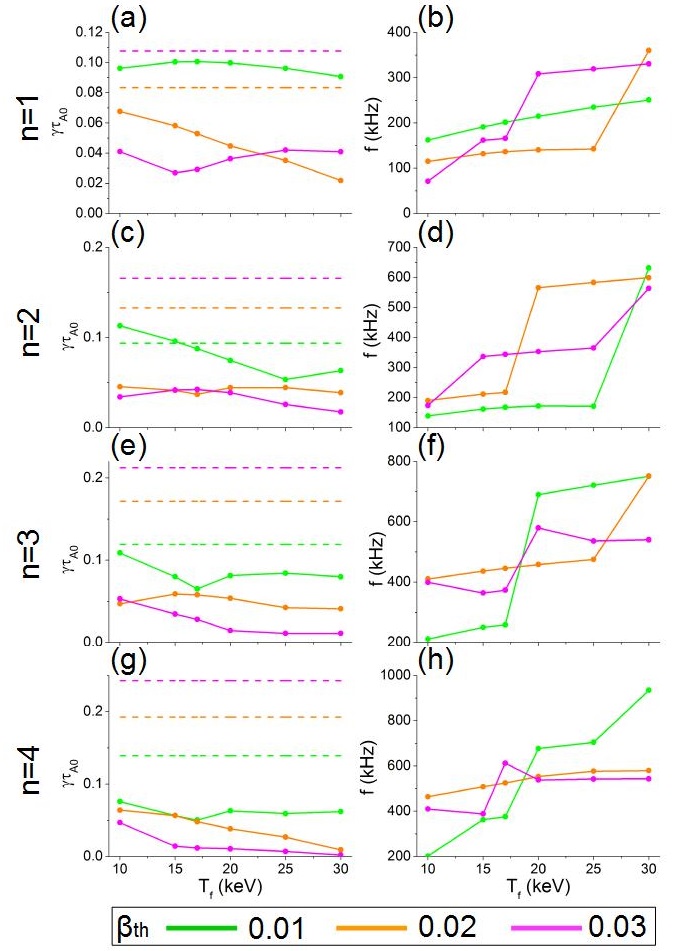} 
\caption{Growth rate and frequency of the $n=1$ to $4$ modes for different EP energies if $\beta_{th} = 0.01$, $0.02$ and $0.03$. EP $\beta=0.01$ if $\beta_{th} = 0.01$, EP $\beta=0.005$ if $\beta_{th} = 0.02$ and EP $\beta=0.0033$ if $\beta_{th} = 0.03$. The dashed lines indicate the growth rate of the $n=1$ to $4$ pressure gradient driven modes.}
\label{FIG:14}
\end{figure}

\begin{figure}[h]
\centering
\includegraphics[width=0.45\textwidth]{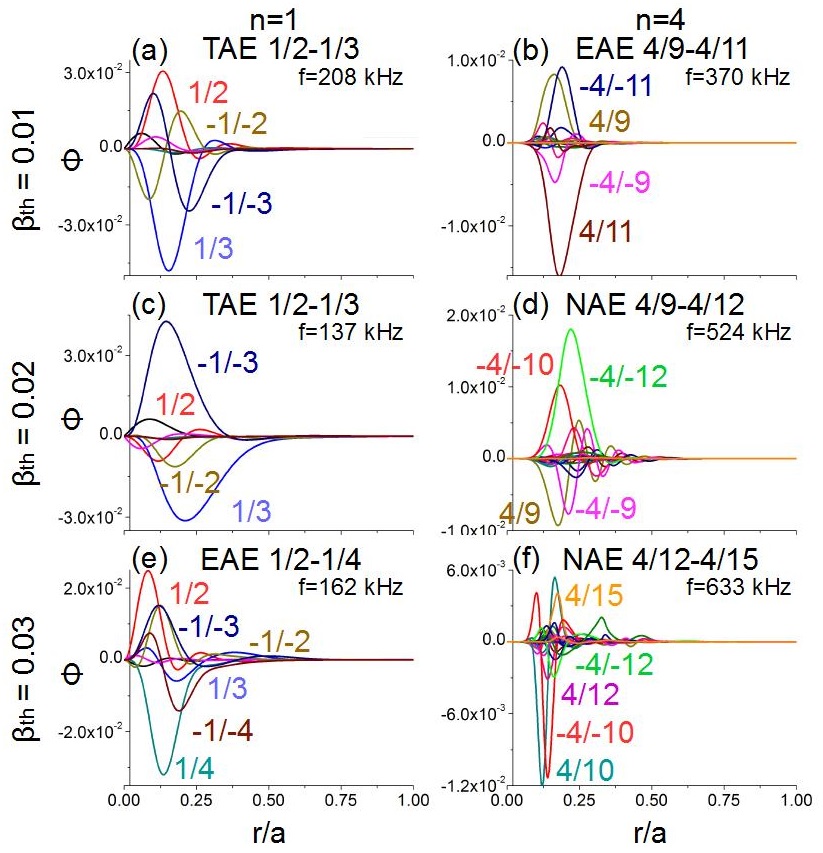} 
\caption{Eigenfunction of the $n=1$ AE if EP $\beta = 0.005$ and $T_{f} = 17$ keV if (a) $\beta_{th} = 0.01$, (c) $\beta_{th} = 0.02$ and (e) $\beta_{th} = 0.03$. Eigenfunction of the $n=4$ AE if EP $\beta = 0.005$ and $T_{f} = 17$ keV if (b) $\beta_{th} = 0.01$, (d) $\beta_{th} = 0.02$ and (f) $\beta_{th} = 0.03$.}
\label{FIG:15}
\end{figure}

Figure~\ref{FIG:14} shows the growth rate and frequency of the $n=1$ to $4$ AEs for different EP energies if $\beta_{th} = 0.01$, $0.02$ and $0.03$. In the simulations the relation $\beta_{f} \propto 1 / \beta_{th}$ is assumed, thus given that the EP $\beta=0.01$ in the simulations with $\beta_{th} = 0.01$, the EP $\beta$ is $0.005$ in the simulations with $\beta_{th} = 0.02$ and $0.0033$ in the simulations with $\beta_{th} = 0.03$. The growth rate of the $n=1$ to $4$ AE decreases if the $\beta_{th}$ increases (panel a, c, e and g). The growth rate of the $n=1$ to $4$ AEs decreases as the EP energy increases, thus EP at the end of the slowing down process show the largest growth rates. In addition, weakly thermalized EP mainly destabilize high frequency AEs (EAE and NAE, panels b, d, f and h). Figure~\ref{FIG:15} shows the eigenfunction of the $n=1$ and $n=4$ AEs if the EP $\beta = 0.005$ and the $T_{f} = 17$ keV for different $\beta_{th}$. The large modification of the continuum gaps and the iota profile as the $\beta_{th}$ increases leads to different eigenfunction structures. If $\beta_{th}=0.01$ a $1/2-1/3$ TAE with $208$ kHz and a $4/9-4/11$ EAE with $370$ kHz are unstable, although if the $\beta_{th}$ increases the frequency range of the different continuum gaps decreases thus a $1/2-1/3$ TAE with $137$ kHz and a $4/9-4/12$ EAE with $524$ kHz are triggered if $\beta_{th}=0.02$, as well as a $1/2-1/4$ EAE with $162$ kHz and a $4/12-4/15$ EAE with $633$ kHz are triggered if $\beta_{th}=0.03$.

\begin{figure}[h]
\centering
\includegraphics[width=0.45\textwidth]{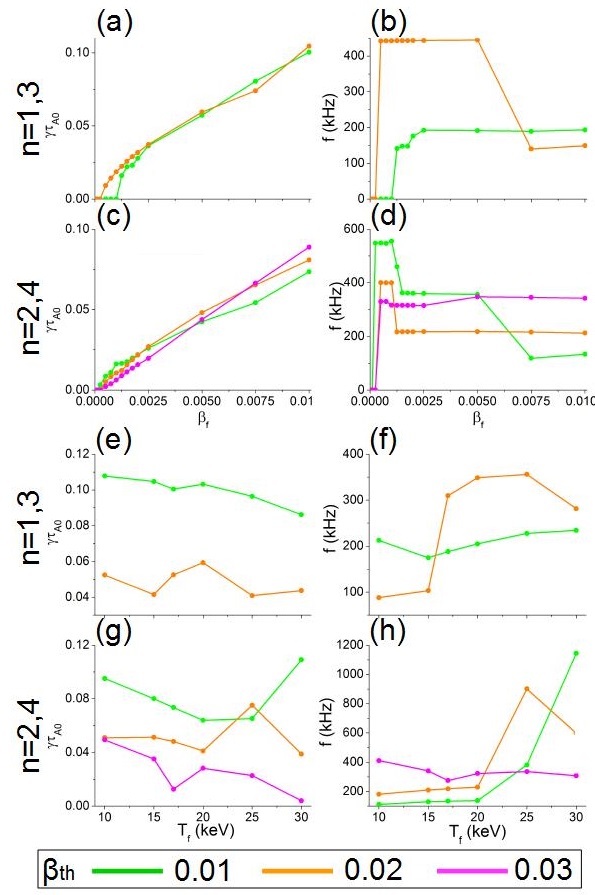} 
\caption{Growth rate and frequency of the $n=1,3$ and $2,4$ AEs for different EP $\beta$ ($T_{f}=17$ keV) if $\beta_{th}=0.01$, $0.02$ and $0.03$. Growth rate and frequency of the $n=1,3$ and $2,4$ AEs for different EP energies if $\beta_{th}=0.01$, $0.02$ and $0.03$, with EP $\beta=0.01$ if $\beta_{th} = 0.01$, EP $\beta=0.005$ if $\beta_{th} = 0.02$ and EP $\beta=0.0033$ if $\beta_{th} = 0.03$.}
\label{FIG:16}
\end{figure}

Figure~\ref{FIG:16}, panels a to d, show the growth rate and frequency of the $n=1,3$ and $2,4$ helical families in simulations with different EP $\beta$ fixed $T_{f}=17$ keV if $\beta_{th}=0.01$, $0.02$ and $0.03$. The $n=1,3$ AEs are stable if $\beta_{th} = 0.03$ and the growth rate is similar in the simulations with $\beta_{th} = 0.01$ and $0.02$. The simulations with  $\beta_{th} = 0.01$ indicate the destabilization of $n=1,3$ AEs with $f = 140$ kHz, although in the simulations with $\beta_{th} = 0.02$ $n=1,3$ AEs with $f = 442$ kHz are unstable, showing a transition to a lower frequency AE family with $f = 140$ kHz as the EP $\beta$ increases. The $n=2,4$ AEs are unstable in the frequency range of the EAEs and NAEs showing a similar growth rate for all the $\beta_{th}$ analyzed, except if $\beta_{th} = 0.01$ above a given EP $\beta$ , showing the destabilization of an AE with $f = 140$ kHz, the TAE frequency range. The growth rate of the $n=1,3$ and $2,4$ AE decrease with the $\beta_{th}$ and EP energy, with the exception of a local increase of the growth rate if there is a transition to a higher frequency AE family. If the simulations with and without helical couplings are compared (see fig~\ref{FIG:13} and~\ref{FIG:14} against fig~\ref{FIG:16}), the growth rate and frequency of the AEs destabilized by the $n=1,3$ helical family and the $n=1$ toroidal family for different $\beta_{th}$, EP $\beta$ and $T_{f}$ are very similar, as well as the AEs triggered by the $n=2,4$ helical family regarding the $n=2$ toroidal family. Consequently, the effect of the helical couplings on the AE stability is weak; this is the reason why the HAEs are not dominant in the finite $\beta_{th}$ cases. One possible explanation is the narrower HAE gaps in the finite $\beta_{th}$ cases with respect to the vacuum case.

\section{Effect of the thermal ion FLR, EP FLR and e-i Landau damping on the AE stability \label{sec:damping}}

The thermal ion FLR, EP FLR and e-i Landau damping effects reduce the available free energy required to destabilized the AEs. Thus, if these damping terms are included in the simulations the growth rate of the AEs decreases or the modes can be stable. In addition, these damping terms have a different effect with respect to the frequency of the AE, leading to a transition between dominant modes of different AE families. The numerical model includes the FLR damping effect for the EP and thermal ions, that is to say, averaging of the instability produced fields by the circular motion around the magnetic field lines and the energy lost as radiation by the thermal ions. In addition, the e-i Landau damping transfers energy from the AE to the thermal electron and ions. In the simulations the Larmor radius (defined as $r_{L} = m v_{\perp} / q B$ with $v_{\perp}$ the particle velocity component perpendicular to the magnetic field) of the EP is $0.005$ and $0.0006$ m for the thermal ions, $0.2$ and $0.025$ times the width of the AE eigenfunction, respectively. For a further description of the implementation of the damping effects please see the Appendix.

\begin{figure}[h]
\centering
\includegraphics[width=0.45\textwidth]{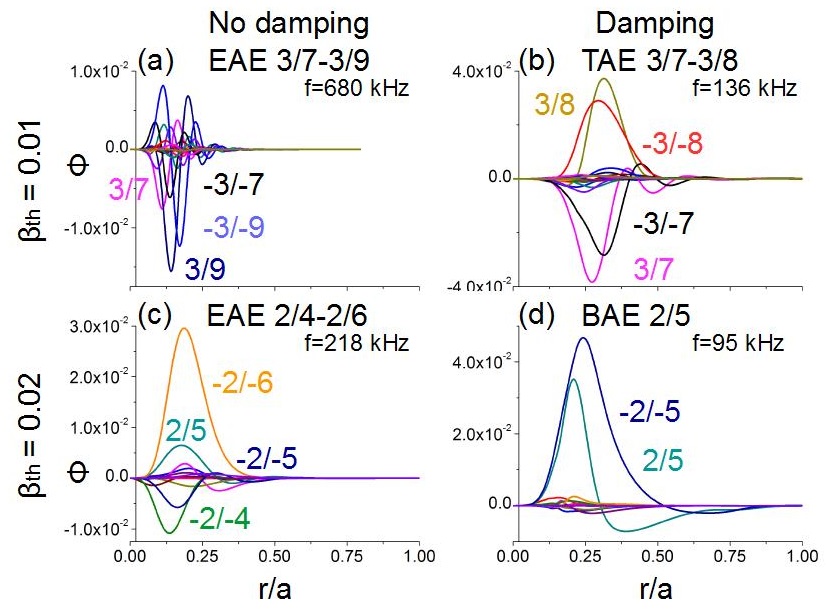} 
\caption{Eigenfunction of the $n=3$ AE if EP $\beta = 0.005$, $T_{f} = 17$ keV and $\beta_{th} = 0.01$ if (a) without damping effects and (b) with damping effects. Eigenfunction of the $n=2$ AE if EP $\beta = 0.005$, $T_{f} = 17$ keV and $\beta_{th} = 0.02$ if (c) without damping effects and (d) with damping effects.}
\label{FIG:17}
\end{figure}

Table~\ref{Table:2} compares the growth rate and frequency of the AEs in simulations with and without damping effects for EP $\beta = 0.005$, $T_{f} = 17$ keV and different thermal $\beta$ values. The simulations with damping effects show a decrease of the AE growth rate between a $20$ to $80 \%$. The largest decrease of the growth rate is observed for modes in the frequency range of the EAE and NAE gaps, because the FLR effects are enhanced as the scale length of the AEs decreases. In addition, the $n=4$ AE are stable if the $\beta_{th} = 0.01$, the $n=3$ and $4$ AEs if the $\beta_{th} = 0.02$ although if the $\beta_{th} = 0.03$ the $n=3$, $n=4$ and both helical AE families are stable, a result consistent with the improved AE stability calculated for high $\beta$ CFQS operation scenarios. Several simulations show a transition between dominant modes of high frequency to low frequency AE families in the simulations with damping effects. An example of such transition is observed in figure~\ref{FIG:17} showing the eigenfunction of $n=2$ and $n=3$ AEs in simulations with and without damping effects. The simulation without damping effects and $\beta_{th} = 0.01$ shows the destabilization of a $3/7-3/9$ EAE with $680$ kHz (panel a) although if the damping effects are included the dominant mode is a $3/7-3/8$ TAE with $136$ kHz (panel b). The same way, the simulation without damping effects and $\beta_{th} = 0.02$ shows the destabilization of a $2/4-2/6$ EAE with $218$ kHz (panel c) although if the damping effects are included the dominant mode is a $2/5$ BAE with $95$ kHz (panel d). The transition between dominant modes from high frequency to lower frequency AE families is caused by the stronger damping of EAE/NAEs relative to BAE/TAEs.

\begin{table}[h]
\centering
\begin{tabular}{c}
$\beta_{th} = 0.01$  \\
\end{tabular}

\begin{tabular}{c  c  c}
  & No damping & Damping  \\
\end{tabular}
\centering
\begin{tabular}{c | c  c | c c}
\hline
n  & $\gamma\tau_{0}$ & f (kHz) & $\gamma\tau_{0}$ & f (kHz) \\ \hline
$1$ & $0.06$ & $208$ & $0.04$ & $110$  \\
$2$ & $0.04$ & $126$ & $0.02$ & $44$  \\
$3$ & $0.05$ & $680$ & $0.01$ & $136$  \\
$4$ & $0.03$ & $370$ & \textcolor{blue}{stb.} & \textcolor{blue}{stb.}  \\
$1,3$ & $0.06$ & $189$ & $0.03$ & $99$  \\
$2,4$ & $0.04$ & $356$ & $0.02$ & $34$  \\ \hline
\end{tabular}

\centering
\begin{tabular}{c}
$\beta_{th} = 0.02$  \\
\end{tabular}

\begin{tabular}{c  c  c}
  & No damping & Damping  \\
\end{tabular}
\centering
\begin{tabular}{c | c  c | c c}
\hline
n  & $\gamma\tau_{0}$ & f (kHz) & $\gamma\tau_{0}$ & f (kHz) \\ \hline
$1$ & $0.05$ & $137$ & $0.04$ & $109$  \\
$2$ & $0.04$ & $218$ & $0.01$ & $95$  \\
$3$ & $0.06$ & $446$ & \textcolor{blue}{stb.} & \textcolor{blue}{stb.}  \\
$4$ & $0.05$ & $524$ & \textcolor{blue}{stb.} & \textcolor{blue}{stb.}  \\
$1,3$ & $0.06$ & $140$ & $0.02$ & $88$  \\
$2,4$ & $0.05$ & $218$ & $0.01$ & $174$  \\ \hline
\end{tabular}

\centering
\begin{tabular}{c}
$\beta_{th} = 0.03$  \\
\end{tabular}

\begin{tabular}{c  c  c}
  & No damping & Damping  \\
\end{tabular}
\centering
\begin{tabular}{c | c  c | c c}
\hline
n  & $\gamma\tau_{0}$ & f (kHz) & $\gamma\tau_{0}$ & f (kHz) \\ \hline
$1$ & $0.04$ & $162$ & $0.03$ & $70$  \\
$2$ & $0.06$ & $345$ & $0.01$ & $328$  \\
$3$ & $0.04$ & $377$ & \textcolor{blue}{stb.} & \textcolor{blue}{stb.}  \\
$4$ & $0.02$ & $633$ & \textcolor{blue}{stb.} & \textcolor{blue}{stb.}  \\
$1,3$ & \textcolor{blue}{stb.} & \textcolor{blue}{stb.} & \textcolor{blue}{stb.} & \textcolor{blue}{stb.}  \\
$2,4$ & $0.04$ & $348$ & \textcolor{blue}{stb.} & \textcolor{blue}{stb.}  \\ \hline
\end{tabular}

\caption{Growth rate and frequency of the AEs in simulations with and without damping effects for EP $\beta = 0.005$, $T_{f} = 17$ keV and thermal $\beta= 0.01$, $0.02$ and $0.03$.} \label{Table:2}
\end{table}

In summary, including the FLR and e-i Landau damping effects in the simulations leads to the stabilization of the $n=1$ to $4$ EAE/NAEs triggered in the simulations without damping effects (except the $n=2$ NAE with $f=328$ kHz in the simulation with $\beta_{th} = 0.03$), as well as a decrease of the growth rate and frequency of the dominant $n=1$ to $n=4$ BAE/TAEs. In addition, the results of the simulations with damping effects reinforce the AE stability optimization trend regarding the CFQS high $\beta$ operation scenarios.

\section{Conclusions and optimization trends \label{sec:conclusions}}

The analysis of the simulations performed by FAR3d code indicates the destabilization of $n=1$ to $4$ AEs as well as $n=2,4$ HAEs in NBI heated CFQS plasma. The EP $\beta$ threshold to destabilize AEs is calculated for EP during the slowing down process, that is to say, from weakly thermalized EP ($T_{f} \ge 25$ keV) to EP at the last stage of the slowing down process ($T_{f} \le 15$ keV). 

The study of the AE destabilization threshold with respect to the EP $\beta$ for different EP energies indicates that, an optimized operational regime of the NBI requires a power injection not exceeding EP $\beta = 0.001$ to avoid the destabilization of high frequency AEs (EAE and NAE) by weakly thermalized EP. In addition, TAEs  and BAEs that can be triggered by EPs during the slowing down process ($T_{f} \le 17$ keV) are stable if EP $\beta < 0.0005$.

The analysis of off-axis NBI models shows no evident optimization trend with respect to an on-axis NBI injection, because the growth rate of the $n=1$ AE decreases although the growth rate of the $n=2-4$ AE increase. On the other hand, there is a transition to lower frequency AEs if the beam is deposited in the middle plasma or the periphery. It should be noted that the effect of the Shafranov shift leads to the outward displacement of the magnetic axis as $\beta_{th}$ increases, thus the NBI deposition region changes from on-axis in low $\beta$ operations to off-axis in high $\beta$ operations. Consequently, a detailed analysis of the AE stability for off-axis NBI depositions is required in high $\beta$ operations.

The simulations performed adding the effect of the helical couplings indicate the possible destabilization of $2/5$ and $2/6$ HAEs by EP with $T_{f} \ge 15$ keV if EP $\beta \ge 0.0005$. On the other hand, the simulations show stable $n=1,3$ HAEs. The analysis of the Alfven gaps indicates narrow $n=1,3$ HAE gaps with respect to the $n=2,4$ HAE gaps, thus the continuum damping is strong enough to stabilize the $n=1,3$ HAE.

The analysis of the AE stability in models with a finite $\beta_{th}$ indicates that CFQS operation with high $\beta$ should have an improved AE stability with respect to low $\beta$ discharges. High $\beta$ operations show slender Alfven gaps, thus the continuum damping is enhanced increasing the EP $\beta$ threshold required to destabilized the AEs. The simulations show a decrease of the growth rate of the $n=1$ to $4$ AEs as well as the $n=1,3$ and $2,4$ helical families as $\beta_{th}$ increases. It should be noted that the model assumes an increase of $\beta_{th}$ caused by a larger thermal plasma density, thus the AE stability trend can be different if the $\beta_{th}$ increases by a larger thermal plasma temperature. Future analysis will be dedicated to clarify the effect of the thermal plasma temperature on the AE stability in CFQS plasma.

The simulations including the effect of the thermal ion FLR, EP FLR and e-i Landau damping indicate the stabilization of the $n=1$ to $4$ EAE/NAEs triggered in the simulations without damping effects. Also, the $n=1$ to $4$ BAE/TAEs show a lower growth rate and frequency relative to the simulations without damping effects, as well as the improved AE stability of CFQS operational scenarios with a high thermal $\beta$.

In summary, the heating efficiency of CFQS plasma heated by a tangential NBI can decrease due to the destabilization of $n=1$ to $4$ BAE/TAEs and $n=2,4$ HAEs above a given injection intensity threshold, particularly if the thermal $\beta$ of the discharge is low. Nevertheless, the heating efficiency can be improved by an optimized NBI operational regime with respect to the NBI voltage, injection intensity and deposition region as well as the thermal plasma parameters. The present analysis will be extended to Stellarators that explores different quasi-symmetries, identifying the magnetic configurations that show an optimal AE stability, particularly for reactor-relevant plasma.

\section*{Appendix}

The model approximations used to describe the EP destabilizing effect is discussed in this appendix.

\subsection*{EP distribution function}

The EP distribution in the simulations is a Maxwellian which has the same second moment, the effective EP temperature, as the slowing down distribution:
\begin{equation}
f_{SD} = \frac{S_{0} \tau_{s}}{4\pi} \frac{1}{v^{3} + v^{3}_{c}} H(v - v_{EP,NBI})
\end{equation}
The Maxwellian distribution is:
\begin{equation}
\label{eq:Max}
f_{Max} = N_{M}e^{\frac{-mv^{2}}{K_{B}T}}
\end{equation}
where $v_{c} = (3\sqrt{\pi} m_{e}/4m_{i})^{1/3} \cdot v_{e}$, with $m_{e}$ the electron mass, $m_{i}$ the ion mass and $v_{e}$ the electron velocity, and $v_{EP,NBI} = \sqrt{2E_{EP,NBI}/m_{EP,NBI}}$ the beam particle birth velocity with $E_{EP,NBI}$ the initial beam particle energy and $m_{EP,NBI}$ the beam particles mass. $\tau_{s}$ is the slowing down time.
The EP model cannot reproduce the destabilization caused by anisotropic beams or ICRF driven EP, although the destabilizing effect of passing particles generated by a tangential NBI  is reproduced.
The averaged square velocity of the slowing down and Maxwellian distribution is selected to be the same ($\langle v^{2} \rangle_{Max} = \langle v^{2} \rangle_{SD}$), where the averaged square velocity is defined as:
\begin{equation}
\langle v^{2} \rangle = \frac{\int fv^{2}dv^{3}}{\int fdv^{3}}
\end{equation}
The assumption of the model is that the averaged Maxwellian is similar to the thermalized velocity of the EP, thus:
\begin{eqnarray} 
\langle v^{2} \rangle_{Max} = \frac{\left(\frac{K_{B}T_{f}}{m_{f}}\right)^{5/2} \int^{\infty}_{0} e^{-x^{2}}x^{4}dx}{\left(\frac{K_{B}T_{f}}{m_{f}}\right)^{3/2} \int^{\infty}_{0} e^{-x^{2}}x^{2}dx} \nonumber\\
\hspace{1.5cm} \approx \frac{K_{B}T_{f}}{m_{f}} \approx v_{th,f}^{2} 
\end{eqnarray} 
with $x^{2} = \frac{m_{f} v^{2}}{K_{B}T_{f}}$, where:
\begin{equation}
\langle v^{2} \rangle_{SD} = \frac{\int^{v_{EP,NBI}}_{0} \frac{v^{4}dv}{v^{3} + v^{3}_{c}}} {{\int^{v_{EP,NBI}}_{0} \frac{v^{2}dv}{v^{3} + v^{3}_{c}}}} = v^2_{c} \frac{\int^{v_{EP,NBI}/v_{c}}_{0} \frac{x^{4}dx}{x^{3} + 1}} {{\int^{v_{EP,NBI}/v_{c}}_{0} \frac{x^{2}dx}{x^{3} + 1}}}
\end{equation}
with $x = v / v_{c}$. Consequently, if the electron temperature is $1$ keV:
\begin{equation}
T_{f} = 0.573 E_{NBI}
\end{equation}
A single Maxwellian distribution function cannot reproduce the energy transfer generated by the resonances of a slowing down distribution function, because the energy exchange depends on the phase-space gradient. Nevertheless, the parametric studies performed regarding the EP density and energy, NBI deposition region and thermal plasma parameters provide useful information for future optimization studies. Consequently, FAR3d results may be verified by models including an anisotropic slowing down distribution function.

\section*{Appendix B}

The implementation of the FLR and electron-ion Landau damping effects in the numerical model is described in this section. For further information please see \cite{40}.

\subsection*{Finite Larmor Radius effects}

The contribution of the thermal ion FLR damping effect is included in the equations of the poloidal flux ($\psi$) and the vorticity ($U$):

$$\frac{\partial \psi}{\partial t} = ... + \rho_{i}^{2} \sqrt{\frac{\pi}{2}} \frac{v_{A}^{2}}{v_{Te}}|\nabla_{||}|\nabla_{\perp}^2 \psi $$
$$\frac{\partial U}{\partial t} = ... + \omega_{r}  \rho_{i}^{2} \nabla_{\perp}^2 U  $$
with $\rho_{i}$ the Larmor radius of the thermal ions normalized to the minor radius, the Alfven velocity ($v_{A}$) and thermal velocity ($v_{Te}$) normalized to the Alfven velocity at the magnetic axis and the $\omega_{r}$ the target AE frequency normalized to the Alfven time. A Pade approximation is used. This is implemented by introducing the variable $Q$ as well as the auxiliary equation:
$$0 = Q - \nabla_{\perp}^2 \psi$$

The contribution of the EP FLR damping effect is included in the equations of the EP density ($n_{f}$) and parallel velocity ($v_{||,f}$):

$$\frac{\partial n_{f}}{\partial t} = ... + \epsilon^2  \omega_{r} \Omega_{cf} \frac{n_{f0}}{v_{th,f}^{2}}W - n_{f0} \Omega_{*}(W)$$
with $\epsilon$ the aspect ratio, $\Omega_{cf}$ the normalized cyclotron frequency to the Alfven time, $W$ the auxiliary variable introduced by the Pade approximation, $\Omega_{*}$ the operator that models the diamagnetic drift frequency of the EP and $v_{th,f}$ the thermalized velocity of the EP. Thus, a new equation is included in the numerical model:
$$0 = (1-\rho_{f}^{2} \nabla_{\perp}^{2}) W + \rho_{f}^{2} \nabla_{\perp}^{2} \Phi$$
with $\Phi$ a stream function proportional to the electrostatic potential. The EP FLR term in the parallel velocity moment equation is:
$$\frac{\partial v_{||,f}}{\partial t} = ... +  v_{th,f}^{2} \frac{1}{J-\rlap{-} \iota I} \frac{1}{n_{f0}} \frac{1}{\rho} \frac{d n_{f0}}{d \rho}(IX_{1}-JX_{2})$$
with $J$ the poloidal current, $I$ the toroidal current, $\rlap{-} \iota$ the rotational transform and $X_{1},X_{2}$ the auxiliary variables given by the Pade approximation. Thus, two new equations are added to the numerical model:
$$(1-\rho_{f}^{2} \nabla_{\perp}^{2}) X_{1} - \frac{\partial \psi}{\partial \zeta} = 0$$
$$(1-\rho_{f}^{2} \nabla_{\perp}^{2}) X_{2} - \frac{\partial \psi}{\partial \theta} = 0$$

\subsection*{Electron-ion Landau damping effects}

The contribution of the e-i damping effect is included in the vorticity equation through a term of the form:

$$\frac{\partial U}{\partial t} = ... - \frac{\beta_{0i}}{2\epsilon^2 \omega_{r}}\frac{T_{i}}{T_{e}}p_{i,eq}(S_{ei})_{imag}\Omega_{d}^2(\Phi)$$
with $\beta_{0i}$ the thermal ion $\beta$ at the magnetic axis, $T_{i} / T_{e}$ the ratio of the ion and electron temperature, $p_{i,eq}$ the equilibrium pressure of the thermal ions, $(S_{ei})_{imag}$ the imaginary component of the e-i damping term and $\Omega_{d}$ the operator that models the average drift velocity of the EP. The complete definition of $(S_{ei})_{imag}$ is written in \cite{40}.

\ack
The authors would like to thank CFQS and LHD technical staff for their contributions. This work was supported by NIFS07KLPH004 and the project 2019-T1/AMB-13648 founded by the Comunidad de Madrid. Data available on request from the authors. The authors also want to acknowledge K. Y. Watanabe for fruitful discussion.

\hfill \break

\end{document}